\documentclass[12pt]{article}
\usepackage{epsfig,amssymb,psfrag,amsmath}

\catcode`\@=11
\def \thesection {\arabic{section}.}

\def \be  {\begin{equation}}
\def \ee  {\end{equation}}
\def \ba  {\begin{eqnarray}}
\def \ea  {\end{eqnarray}}
\def \baa {\begin{eqnarray*}}
\def \eaa {\end{eqnarray*}}
\def \bb  {\begin {thebibliography} }
\def \eb  {\end{thebibliography}}
\def \lab #1 {\label{#1}}

\newcommand\re[1]{(\ref{#1})}

\def \matrix #1 {\left(\begin{array}{cc} #1 \end{array}\right)}

\def \tr {\mathop{\rm tr}\nolimits}

\newcommand\lr[1]{{\left({#1}\right)}}

\newcommand \vev [1] {\langle{#1}\rangle}

\newcommand{\as}{\ifmmode\alpha_{\rm s}\else{$\alpha_{\rm s}$}\fi}
\newcommand{\asbar}{\ifmmode\bar{\alpha}_{\rm s}\else{$\bar{\alpha}_{\rm s}$}\fi}

\newcommand{\ft}[2]{{\textstyle\frac{#1}{#2}}}

\font\cmss=cmss12 
\def\inbar{\,\vrule height1.5ex width.4pt depth0pt}
\def\IC{\relax\hbox{$\inbar\kern-.3em{\rm C}$}}
\def\IZ{\relax{\hbox{\cmss Z\kern-.4em Z}}}
\def\IR{{\hbox{{\rm I}\kern-.2em\hbox{\rm R}}}}

\def\IP{{\hbox{{\rm I}\kern-.2em\hbox{\rm P}}}}
\def\II{\hbox{{1}\kern-.25em\hbox{l}}}

\def\numberbysection{\@addtoreset{equation}{section}
                     \def\theequation{\thesection\arabic{equation}}}
\numberbysection

\newbox\lett\newdimen\lheight\newdimen\lwidth
\def\ontop#1#2{\setbox\lett=\hbox{#2}\lheight\ht\lett
\multiply\lheight by 12 \divide\lheight by 10\relax%
\lwidth\wd\lett \multiply\lwidth by 8 \divide\lwidth by 10\relax #2\kern-\lwidth%
\raise\lheight\hbox{{$\scriptstyle #1$}}\kern.1ex}

\begin{document}

\begin{titlepage}
\begin{flushright}
\begin{tabular}{l}
LPT--Orsay--04--18 \\
RUB--TP2--01/04 \\
DOE/ER/40762--303 \\
UMD-PP\#04--025 \\
hep-th/0403085
\end{tabular}
\end{flushright}

\vskip1cm

\centerline{\large \bf Quantum integrability in (super) Yang-Mills theory on
the light-cone}

\vspace{1cm}

\centerline{\sc A.V. Belitsky$^a$, S.\'E. Derkachov$^{b,c}$,
                G.P. Korchemsky$^b$, A.N. Manashov$^{d,c}$}

\vspace{10mm}

\centerline{\it $^a$Department of Physics, University of Maryland
at College Park} \centerline{\it College Park, MD 20742-4111, USA}

\vspace{3mm}

\centerline{\it $^b$Laboratoire de Physique
Th\'eorique\footnote{Unit\'e
                    Mixte de Recherche du CNRS (UMR 8627).},
                    Universit\'e de Paris XI}
\centerline{\it 91405 Orsay C\'edex, France}

\vspace{3mm}

\centerline{\it $^c$Department of Theoretical Physics,  St.-Petersburg State
University}

\centerline{\it 199034, St.-Petersburg, Russia}

\vspace{3mm}

\centerline{\it $^d$Institut f\"ur Theoretische Physik II,
                    Ruhr-Universit\"at Bochum}
\centerline{\it 44780 Bochum, Germany}

\def\thefootnote{\fnsymbol{footnote}}%
\vspace{1cm}

\centerline{\bf Abstract}

\vspace{5mm}

We employ the light-cone formalism to construct in the (super) Yang-Mills
theories in the multi-color limit the one-loop dilatation operator acting on
single trace products of chiral superfields separated by light-like distances. In
the $\mathcal{N}=4$ Yang-Mills theory it exhausts all Wilson operators of the
maximal Lorentz spin while in nonsupersymmetric Yang-Mills theory it is
restricted to the sector of maximal helicity gluonic operators. We show that the
dilatation operator in all ${\cal N}-$extended super Yang-Mills theories is given
by the same integral operator which acts on the $(\mathcal{N}+1)-$dimensional
superspace and is invariant under the $SL(2|\mathcal{N})$ superconformal
transformations. We construct the $R-$matrix on this space and identify the
dilatation operator as the Hamiltonian of the Heisenberg $SL(2|{\cal N})$ spin
chain.

\end{titlepage}

\setcounter{footnote} 0

\section{Introduction}

It has been recently recognized that four-dimensional Yang-Mills theory
possesses a hidden symmetry. It is not manifest in the classical Lagrangian and
rather reveals itself on the quantum level through integrability properties of
evolution equations governing the energy dependence of scattering amplitudes in
the Regge limit \cite{Lip94,FadKor94} and renormalization group equations for
composite Wilson operators \cite{BraDerMan98,BraDerKorMan99,Bel99,DerKorMan00}.
In both cases, the evolution equations can be brought to the form of a
Schr\"odinger equation which contains a large enough number of hidden integrals
of motion to make it completely integrable. The underlying integrable
structures have been identified as the celebrated Heisenberg spin magnet and
its generalization. Their stringy interpretation has been proposed
in~\cite{BelGorKor03}.

In this paper, we shall study renormalization of Wilson operators in (super)
Yang-Mills theories. These operators mix under renormalization and the
corresponding mixing matrix defines a representation of the dilatation
operator~\cite{Ohrndorf}. The integrability of the latter has been discovered
in QCD in the sector of gauge-invariant Wilson operators of the maximal
helicity, the so-called quasipartonic operators~\cite{BFKL}. To one-loop order,
the dilatation operator inherits the conformal symmetry of the classical QCD
Lagrangian and for the Wilson operators of the maximal Lorentz spin the full
conformal group is reduced to the $SL(2)$ subgroup. In the multi-color limit,
the mixing problem for such operators can be reformulated as an eigenvalue
problem for a one-dimensional chain of $SL(2)$ conformal spins resided on sites
of original QCD fields. Depending on the field content of Wilson operators, two
integrable structures have been revealed: closed spin chains for multi-quark
and multi-gluon operators, and open spin chains for operators involving
fundamental matter at the ends of an array of gluons.

Similar integrability phenomena have been recently found in the context of
maximally supersymmetric $\mathcal{N}=4$ super-Yang-Mills (SYM) theory. There,
the one-loop mixing matrix in the sector of scalar operators has been identified
in the multi-color limit as a Hamiltonian of the Heisenberg $SO(6)$ spin
chain~\cite{MinZar02,BeiKriSta03} and its generalization to arbitrary Wilson
operators led to the identification of the $\mathcal{N}=4$ dilatation operator
with the Hamiltonian of the $PSU(2,2|4)$ super-spin chain \cite{BeiSta03}
\footnote{For super Yang-Mills theories with less supersymmetry analogous
analyses have been performed in Refs.~\cite{Stefanski,WanWu03,CheWanWu04}.}.

A natural question arises whether integrability phenomena found in QCD and in the
$\mathcal{N}=4$ SYM theory in the different sectors of Wilson operators are
related to each other. In this paper, we shall argue that integrability is not a
genuine new symmetry of the $\mathcal{N}=4$ SYM theory but is a general feature
of the Yang-Mills theory in the multi-color limit, at least to one-loop order. In
particular, we shall demonstrate how the $\mathcal{N}=4$ integrable structures
are related to those found previously in QCD to which we shall refer as the
$\mathcal{N}=0$ SYM theory. We shall employ the light-cone formalism to construct
in the (super) Yang-Mills theories in the multi-color limit the one-loop
dilatation operator acting on single trace products of chiral superfields
separated by light-like distances. In the $\mathcal{N}=4$ theory it exhausts all
Wilson operators of the maximal Lorentz spin while in $\mathcal{N}=0$ theory it
is restricted to the sector of maximal helicity gluonic operators. The advantage
of this formalism is that the superfields involve only ``physical'', propagating
modes and superconformal transformations can be realized linearly on the product
of superfields ``living'' on the light-cone. This allows one to construct the
one-loop dilatation operator as a quantum mechanical Hamiltonian and map it into
the Heisenberg $SL(2|{\cal N})$ spin chain.

The paper is organized as follows. In Sect.~2 we review the light-cone
formulation of the (super) Yang-Mills theory. In Sect.~3 we define the generating
function for Wilson operators of the maximal Lorentz spin and formulate the
evolution equation governing its scale dependence. The constraints imposed on the
evolution equation by the superconformal symmetry are discussed in Sect.~4. The
one-loop dilatation operator entering this equation is calculated in the
$\mathcal{N}-$extended SYM in Sect.~5. It is identified in Sect.~6 as a
Hamiltonian of the Heisenberg $SL(2|\mathcal{N})$ spin chain. Section 7 contains
concluding remarks.

\section{Yang-Mills theory on the light-cone}
\label{Sect2}

A convenient framework for discussing integrability properties of
(super) Yang-Mills theories is provided by the ``light-cone
formalism''~\cite{KogSop,BriLinNil83,Man83}. In this formalism, one sacrifices
the full manifest covariance of the theory under the Poincar\'e transformations
with advantage of having the possibility to integrate out non-propagating
components of fields and formulate the quantum action in terms of ``physical''
degrees of freedom. Another benefit of using the light-cone formalism in the SYM
theory is that the $\mathcal{N}-$extended supersymmetric algebra is closed
off-shell for the propagating fields and there is no need to introduce auxiliary
fields. This allows one to design an extended superspace formulation of ${\cal N}
= 4$ SYM theory while the covariant form of the same theory does not exist.

In the light-cone formalism, one splits the gauge field into longitudinal and
transverse components $A_\mu^a(x) = (A_+^a, A_-^a, A^a,\bar A^a)$ (with
$A_\pm(x) \equiv \ft1{\sqrt{2}}(A_0(x)\pm A_3(x))$, $A(x) \equiv
\ft1{\sqrt{2}}(A_1(x)+i A_2(x))$ and $\bar A =A^*$) and quantizes the SYM
theory in a noncovariant, light-cone gauge $A^a_+(x) = 0$. Making a similar
decomposition of (Majorana) fermion fields into the so-called ``bad'' and
``good'' components ${\mit\Psi} = {\Pi}_+ {\mit\Psi} + {\Pi}_- {\mit\Psi}
\equiv {\mit\Psi}_+ + {\mit\Psi}_-$ (with $\Pi_\pm = \ft12 \gamma_\pm
\gamma_\mp$), one finds that the fields ${\mit\Psi}_-(x)$ and $A_-(x)$ can be
integrated out in this gauge. The resulting action of SYM theory is expressed
in terms of physical fields---transverse components of the gauge fields, $A(x)$
and $\bar A(x)$, ``good'' components of fermion fields ${\mit\Psi}_+(x)$ and,
in general, scalar fields. Let us summarize the light-cone formulation of SYM
theories by starting with the $\mathcal{N}=4$ model and going down to
$\mathcal{N}=2$, $\mathcal{N}=1$ and, ultimately, $\mathcal{N}=0$ models.

In the $\mathcal{N}=4$ model, the propagating modes are the complex field $A(x)$
describing transverse components of the gauge field, three complex scalar fields
$\phi^{AB}(x)$, complex Grassmann fields $\lambda^A(x)$ defining ``good''
components of four Majorana fermions (with ${\scriptstyle A,B} =1,\ldots,4$),
and conjugated to them are fields $\bar A(x)$, $\bar \phi_{AB}= \left( \phi^{AB}
\right)^* = \ft12\varepsilon_{ABCD} \phi^{CD}$ and $\bar\lambda_A(x)$. It is
tacitly assumed that the fields belong to the adjoint representation of the
$SU(N_c)$ group. In the light-cone formalism, all propagating fields can be
combined into a single scalar superfield ${\mit\Phi}(x_\mu, \theta^A,
\bar\theta_A)$, and the action of the $\mathcal{N} = 4$ SYM reads
as~\cite{BriLinNil83}
\ba
S_{\mathcal{N}=4}
\!\!\!&=&\!\!\!
\int d^4 x\, d^4\theta\, d^4 \bar\theta\,
\Bigg\{
\frac12 \bar{\mit\Phi}^a\frac{\Box}{\partial_+^2}{\mit\Phi}^a
-
\frac23 g f^{abc}
\lr{
\frac{1}{\partial_+}\bar{\mit\Phi}^a{\mit\Phi}^b\bar\partial{\mit\Phi}^c
+
\frac{1}{\partial_+}{\mit\Phi}^a\bar{\mit\Phi}^b\partial\bar{\mit\Phi}^c
}
\nonumber
\\
&&\qquad\qquad\qquad
- \frac12 g^2 f^{abc}f^{ade}
\left(
\frac{1}{\partial_+} ({\mit\Phi}^b\partial_+{\mit\Phi}^c)
\frac{1}{\partial_+} (\bar{\mit\Phi}^d\partial_+\bar{\mit\Phi}^e)
+
\frac12{\mit\Phi}^b\bar{\mit\Phi}^c{\mit\Phi}^d\bar{\mit\Phi}^e
\right)
\Bigg\}
\, ,
\label{N=4-S}
\ea
where $f^{abc}$ are the $SU(N_c)$ structure constants,
$\partial_+=\ft{1}{\sqrt{2}}(\partial_{x_0}-\partial_{x_3})$,
$\partial=\ft{1}{\sqrt{2}}(\partial_{x_1}+i\partial_{x_2})$, $\bar
\partial= (\partial)^*$ and the integration measure over Grassmann
variables is normalized as $\int d^{\cal N} \theta\, \theta^1 \ldots
\theta^{\cal N} = \int d^{\cal N} \bar\theta \, \bar\theta_1 \ldots
\bar\theta_{\cal N} = 1$. The complex scalar $\mathcal{N}=4$ superfield
is defined as
\begin{eqnarray}
{\mit\Phi} (x, \theta^A, \bar\theta_A)
=
{\rm e}^{\frac12 \bar\theta \cdot \theta \, \partial_+}
\bigg\{ \partial_+^{-1}A(x)
\!\!\!&+&\!\!\!
\theta^A \partial_+^{-1}\bar\lambda_A (x)
+
\frac{i}{2!} \theta^A \theta^B \bar \phi_{AB} (x)
\nonumber\\
&-&\!\!\!
\frac{1}{3!} \varepsilon_{ABCD} \theta^A \theta^B \theta^C \lambda^D (x)
-
\frac{1}{4!} \varepsilon_{ABCD} \theta^A \theta^B \theta^C \theta^D
\partial_+ \bar{A} (x)
\bigg\} . \quad \ \label{N=4-field}
\end{eqnarray}
Here $\bar\theta\cdot \theta = \bar\theta_A \theta^A$ and nonlocal operator
$\partial_+^{-1}$ is defined using the Mandelstam-Leibbrandt
prescription~\cite{Man83}. We recall that the superfield \re{N=4-field} comprises
{\sl all} propagating fields, and expansion in $\theta^A$ can be viewed as an
expansion in different helicity components: $+1$ for $A(x)$, $1/2$ for
$\bar\lambda_A (x)$, $0$ for $\bar\phi_{AB}$, $-1/2$ for $\lambda^A (x)$ and $-1$
for $\bar A(x)$. Notice that the two lowest components in \re{N=4-field} are
nonlocal operators. A unique feature of \re{N=4-field} is that conjugated
superfield is not independent and is related to ${\mit\Phi} (x, \theta^A,
\bar\theta_A)$ as
\be
\bar{\mit\Phi} (x, \theta^A , \bar\theta_A)
=
- \partial_+^{-2} {D_1 D_2 D_3 D_4}
{\mit\Phi} (x, \theta^A, \bar\theta_A)
= - \ft1{4!} \partial_+^{-2} \varepsilon^{ABCD} {D_A D_B D_C D_D}
{\mit\Phi} (x, \theta^A, \bar\theta_A)
\, .
\label{N=4-bar-field}
\ee
Here the notation was introduced for the covariant derivatives in the
superspace
\be
D_A = \partial_{\theta^A} - \ft{1}{2}\bar\theta_A
\partial_+
\, , \qquad
\bar D^A = \partial_{\bar\theta_A}-\ft{1}{2}\theta^A
\partial_+
\, , \qquad
\{ D_A, \bar D^B\} = - \delta_A^B \partial_+
\, .
\ee
The superfields \re{N=4-field} and \re{N=4-bar-field} satisfy the chirality
condition
\be
\bar{D}^A {\mit\Phi} (x , \theta^A , \bar\theta_A)
=
D_A \bar{\mit\Phi} (x, \theta^A , \bar\theta_A)
=
0
\, .
\ee
As was first found in Refs.~\cite{SohWes81,BriLinNil83,HowSteow84,Man83}, all
Green's functions computed from \re{N=4-S} do not contain ultraviolet divergences
to all orders of perturbation theory and, therefore, the $\mathcal{N}=4$
light-cone action \re{N=4-S} defines an ultraviolet finite quantum field theory.
As such, it inherits all symmetries of the classical Lagrangian including the
invariance under superconformal transformations. Later in Sect.~4 we shall make
use of the subgroup of these transformations that leaves the `+'-direction on the
light-cone invariant.

Let us now turn to Yang-Mills theories with less supersymmetry. Their light-cone
formulation can be obtained from $\mathcal{N}=4$ SYM using the ``method of
truncation'' \cite{BriTol83} which is based on the following identity
\be
\int d^4 x \, d^\mathcal{N} \theta \, d^\mathcal{N} \bar\theta \,
\mathcal{L} ({\mit\Phi})
=
(-1)^\mathcal{N}\int d^4 x \,
d^{\mathcal{N}-1} \theta \,
d^{\mathcal{N}-1} \bar\theta \,
\left[
\bar D^\mathcal{N} D_\mathcal{N} \mathcal{L}({\mit\Phi})
\right]
\bigg|_{\theta^{\cal N} = \bar\theta_{\cal N} = 0}
\, .
\label{truncation}
\ee
Applying \re{truncation}, one can rewrite the $\mathcal{N}=4$ model in terms of
one $\mathcal{N}=2$ light-cone Yang-Mills chiral superfield ${\mit\Phi}^{(2)} (x,
\theta^A, \bar\theta_A)$ coupled to the $\mathcal{N}=2$ Wess-Zumino chiral
superfield ${\mit\Psi}^{(2)} (x, \theta^A, \bar\theta_A)$
\be
{\mit\Phi}^{(2)} = {\mit\Phi}^{(4)} (x, \theta^A , \bar\theta_A)
\bigr|_{\theta^3 = \bar\theta_3 = 0 \atop \theta^4 = \bar\theta_4 = 0} \, ,
\qquad {\mit\Psi}^{(2)} = D_3\, {\mit\Phi}^{(4)}(x, \theta^A , \bar\theta_A)
\bigr|_{\theta^3 = \bar\theta_3 = 0 \atop \theta^4 = \bar\theta_4 = 0} \, ,
\label{N=4->2}
\ee
where the superscript refers to the underlying $\mathcal{N}-$extended SYM and
${\mit\Phi}^{(4)}$ is defined in \re{N=4-field}. Putting ${\mit\Psi}^{(2)} = 0$,
one obtains the light-cone formulation of the $\mathcal{N}=2$ SYM
\cite{BriTol83,Smith}
\ba
S_{\mathcal{N}=2}
=
\int d^4 x \, d^2 \theta \, d^2\bar\theta
\bigg\{
\!\!\!&-&\!\!\!
\bar{\mit\Phi}^a{\Box}{\mit\Phi}^a
+
2 g f^{abc}
\lr{
\partial_+ {\mit\Phi}^a \bar{\mit\Phi}^b \bar\partial {\mit\Phi}^c
+
\partial_+ \bar{\mit\Phi}^a {\mit\Phi}^b \partial \bar{\mit\Phi}^c
}
\nonumber \\
&-&\!\!\!
2 g^2 f^{abc}f^{ade} \frac{1}{\partial_+}\lr{\partial_+{\mit\Phi}^b
\bar D^1 \bar D^2 \bar{\mit\Phi}^c}\frac{1}{\partial_+}
\lr{ \partial_+\bar{\mit\Phi}^d D_1 D_2 {\mit\Phi}^e }
\bigg\}
\, . \label{N=2-S}
\ea
Here ${\mit\Phi}\equiv {\mit\Phi}^{(2)}(x, \theta^A , \bar\theta_A)$ is a
complex chiral $\mathcal{N}=2$ superfield. Substituting \re{N=4-field} into
\re{N=4->2} one finds
\be
{\mit\Phi} (x , \theta^A , \bar\theta_A)
= {\rm e}^{\frac12 \bar\theta \cdot \theta \, \partial_+}
\bigg\{
\partial_+^{-1} A(x)
+
\theta^A \partial_+^{-1}\bar\lambda_A (x)
+
\frac{i}{2!} \varepsilon_{AB} \theta^A \theta^B \bar \phi (x)
\bigg\}
\, ,
\label{N=2-field}
\ee
where $\bar \phi \equiv \bar \phi_{12}(x)$ and ${\scriptstyle A,B} = 1,2$.
The conjugated (anti-chiral) superfield $\bar{\mit\Phi} (x, \theta^A,
\bar\theta_A)$ involves the fields $\bar A(x)$, $\lambda^A$ and $\phi$,
and in distinction with the $\mathcal{N=}4$ model, it is independent on
${\mit\Phi}^{(2)}(x, \theta^A , \bar\theta_A)$. In the $\mathcal{N}=2$ light-cone
action \re{N=2-S}, the propagating fields are the transverse components of the
gauge field $A(x)$, one complex scalar field $\phi(x)$ and two complex Grassmann
fields $\lambda^A(x)$ describing ``good'' components of two Majorana fermions.

As a next step, one applies \re{truncation} to truncate the $\mathcal{N}=2$
down to $\mathcal{N}=1$ SYM. Similar to the previous case, one defines two
chiral superfields ${\mit\Phi}^{(1)} = {\mit\Phi}^{(2)} (x, \theta^A ,
\bar\theta_A) |_{\theta^2=\bar\theta_2=0}$ and ${\mit\Psi}^{(1)} = D_2
{\mit\Phi}^{(2)} (x, \theta^A , \bar\theta_A)|_{\theta^2 =
\bar\theta_2 = 0}$ and puts ${\mit\Psi}^{(1)} = 0$ to retain only the
contribution of the $\mathcal{N}=1$ SYM superfield. This leads to
\ba
S_{\mathcal{N}=1}
=
\int d^4 x \, d \theta \, d \bar\theta
\bigg\{
\bar{\mit\Phi}^a{\Box}\, \partial_+ {\mit\Phi}^a
\!\!\!&+&\!\!\!
2 g f^{abc}
\lr{
\partial_+{\mit\Phi}^a \partial_+\bar{\mit\Phi}^b\bar\partial{\mit\Phi}^c
-
\partial_+\bar{\mit\Phi}^a \partial_+{\mit\Phi}^b\partial\bar{\mit\Phi}^c
}
\nonumber\\
&+&\!\!\!
2 g^2 f^{abc}f^{ade} \frac{1}{\partial_+}
\lr{
\partial_+{\mit\Phi}^b
\bar D^1\partial_+\bar{\mit\Phi}^c}\frac{1}{\partial_+}
\lr{\partial_+\bar{\mit\Phi}^d D_1\partial_+{\mit\Phi}^e}
\bigg\}
\, , \ \
\label{N=1-S}
\ea
where the $\mathcal{N}=1$ light-cone chiral superfield ${\mit\Phi} \equiv
{\mit\Phi}^{(1)}(x, \theta, \bar\theta)$ is given by
\be
{\mit\Phi} (x ,  \theta , \bar\theta )
=
{\rm e}^{\frac12  \bar\theta \theta \partial_+}
\bigg\{
\partial_+^{-1}A(x) + \theta\, \partial_+^{-1} \bar\lambda  (x)
\bigg\}
\, ,
\label{N=1-field}
\ee
with $\bar\lambda = \bar\lambda_1(x)$. In the $\mathcal{N}=1$ light-cone
action \re{N=1-S}, the propagating fields are the transverse components of
the gauge fields $A(x)$ and one complex Grassmann field $\lambda(x)$
describing ``good'' component of Majorana fermion.

Finally, we use \re{truncation} to truncate $\mathcal{N}=1$ down to
$\mathcal{N}=0$ Yang-Mills theory. The resulting light-cone action takes
the form
\ba
S_{\mathcal{N}=0}
=
\int d^4 x
\bigg\{
\bar{\mit\Phi}^a{\Box} \, \partial_+^2 {\mit\Phi}^a
\!\!\!&-&\!\!\!
2 g f^{abc}
\lr{
\partial_+{\mit\Phi}^a \partial_+^2 \bar{\mit\Phi}^b \bar\partial {\mit\Phi}^c
+
\partial_+\bar{\mit\Phi}^a \partial_+^2{\mit\Phi}^b\partial\bar{\mit\Phi}^c
}
\nonumber\\
&-&\!\!\!
2 g^2 f^{abc} f^{ade}
\frac{1}{\partial_+}
\lr{
\partial_+ {\mit\Phi}^b
\partial_+^2 \bar{\mit\Phi}^c
}
\frac{1}{\partial_+}
\lr{
\partial_+\bar{\mit\Phi}^d \partial_+^2{\mit\Phi}^e
}
\bigg\}
\, ,
\label{N=0-S}
\ea
where the $\mathcal{N}=0$ field is given by
\be
{\mit\Phi} (x) = {\mit\Phi}^{(1)}(x, \theta , \bar\theta )|_{\theta = \bar\theta = 0}
=
\partial_+^{-1} A(x)
\, .
\label{N=0-field}
\ee
The light-cone action \re{N=0-S} coincides with the well-known expression for the
action of $SU(N_c)$ gluodynamics quantized in the gauge $A_+ (x)=0$~\cite{DK}.

It seems like an overcomplication to work with \re{N=0-field} since one can easily
reformulate the action \re{N=0-S} in terms of a local, field strength tensor
$\partial_+ A(x)$. In a similar manner, the $\mathcal{N}=1$ and $\mathcal{N}=2$
light-cone actions, Eqs.~\re{N=1-S} and \re{N=2-S},  can be formulated in terms
of superfields, ${\mit\Phi}_{\rm new}^{(1)} ={\rm e}^{\frac12
\bar \theta \theta
\partial_+} \{ \bar\lambda (x) + \theta\,
\partial_+ A(x) \}$ and ${\mit\Phi}_{\rm new}^{(2)} = {\rm e}^{\frac12
\bar\theta \cdot \theta \,
\partial_+} \{ i \bar\phi (x) + \theta^A \bar\lambda_A (x) - \frac{1}{2!}
\varepsilon_{AB} \theta^A \theta^B
\partial_+ A(x) \}$, respectively, expressed solely in terms of the elementary
conformal primaries. In the $\mathcal{N}=4$ case, this is not possible on the
basis of simple dimensional considerations alone since the superfield has to
embrace all available particle helicity states. However, in our analysis we
prefer to deal with the superfields involving nonlocal operators. The reason
for this is that as we will show below the evolution operator governing the
scale dependence of the product of such superfields turns out to be universal
for all Yang-Mills theories in question.

\section{Evolution equations on the light-cone}
\label{Sect3}

Let us consider the scale dependence of Wilson operators in the
$\mathcal{N}$-extended SYM. In general, these are local gauge-invariant composite
operators $\mathbb{O}_{\mu_1 \mu_2 \ldots \mu_L}(x)$ built from fundamental
fields -- strength tensor $F_{\mu\nu}(x)$, fermions ${\mit\Psi}^A (x)$, scalars
$\phi^{AB}(x)$ -- and their complex conjugate as well as arbitrary number of
covariant derivatives $D_{\mu} = \partial_\mu - ig \, [A_\mu(x),~ ]$ acting on
them. The Wilson operators carry Lorentz and isotopic $SU(\mathcal{N})$ indices
and can be classified according to irreducible representations of both groups.
Later in this section we shall study the Wilson operators of the maximal Lorentz
spin or, equivalently, the minimal twist. They have the Lorentz structure
completely symmetric and traceless in any pair of Lorentz indices. Such operators
can be obtained by projecting a general operator onto light-like vectors
$\mathbb{O}_L^{\rm (max)}=n^{\mu_1} n^{\mu_2} \ldots n^{\mu_L} \mathbb{O}_{\mu_1
\mu_2 \ldots \mu_L}(x)$ with $n_\mu^2 = 0$.

In the light-cone formalism, the Wilson operators can be built only from
``physical'' components of fundamental fields (transverse components of gauge
fields, ``good'' components of fermions and scalars) and covariant derivatives
acting on them. As was already mentioned, the remaining components are not
dynamically independent and can be expressed (nonlocally) in terms of physical
ones by virtue of the equations of motion. In addition, making use of the
superfield formulation, one can construct the Wilson operators directly from the
light-cone scalar chiral ${\mit\Phi} (x , \theta^A ,
\bar\theta_A)$ and antichiral $\bar{\mit\Phi} (x , \theta^A , \bar\theta_A)$
superfields, covariant derivatives $D_{\mu}$ and derivatives acting on Grassmann
variables $\partial_{\theta^A}$ and $\partial_{\bar\theta_A}$. This construction
holds in the SYM theory regardless the number of supersymmetries involved.
Notice that according to \re{N=4-bar-field} the Wilson operators in the
$\mathcal{N}=4$ SYM can be built solely from chiral superfields.
For $\mathcal{N}\le 2$ the superfields  ${\mit\Phi} (x , \theta^A ,
\bar\theta_A)$ and $\bar{\mit\Phi} (x , \theta^A , \bar\theta_A)$ are
independent on each other and have to be taken into account on equal footing.

In what follows we shall restrict ourselves to the subclass of single trace
maximal spin Wilson operators
built entirely from chiral (or anti-chiral) superfields. In the light-like axial
gauge $n \cdot A \equiv A_+ (x) = 0$, such operators can be constructed from the
strength tensor $n_\mu F^\mu{}_{\nu} = F_{+\nu}(x) = \partial_+ A_\nu (x)$, or
equivalently, $\partial_+ A(x)$ and $\partial_+ \bar A(x)$, fermions ($\lambda^A$
and $\bar\lambda_A$), scalars ($\phi^{AB}$ and $\bar \phi_{AB}$) and covariant
derivatives $n \cdot D = D_+ =
\partial_+$. The generating function for such operators takes the form
\be
\mathbb{O}(Z_1, Z_2, \ldots, Z_L) = \tr \left\{ {\mit\Phi}(x_1 n_\mu,
\theta^A_1 , \bar\theta_{1A}) \, {\mit\Phi}(x_2 n_\mu, \theta^A_2 ,
\bar\theta_{2A}) \ldots {\mit\Phi}(x_L n_\mu, \theta^A_L , \bar\theta_{LA})
\right\} \, . \label{O}
\ee
Here all superfields are located on the light-cone along the `$+$'--direction
defined by a light-like vector $n_\mu$ ($ n_+ = n_\perp = 0$ and $n_- = 1$) and
the variables $x_k$ specify their position. The $Z_k-$variables in the left-hand
side of \re{O} stand for the coordinates of the superfields in the superspace.
According to \re{N=4-field}, a chiral superfield satisfies the relation
${\mit\Phi} (x n_\mu , \theta^A , \bar\theta_A) = {\mit\Phi} (n_\mu(x + \ft12
\bar\theta \cdot\theta ) , \theta^A , 0)$ which allows one to eliminate the
$\bar\theta$-dependence in \re{O} and define the $Z-$coordinates in the
superspace as \footnote{Throughout the paper we adopt the following convention
for the complex conjugation $(\bar\theta \chi)^* = \theta\bar \chi$.}
\be
Z
=
(z, \theta^1, \ldots, \theta^{\mathcal{N}})
\, , \qquad
z \equiv x + \frac12 \bar\theta\cdot \theta
\, .
\ee
Later we shall use a shorthand notation for the chiral light-cone superfield
${\mit\Phi}(Z) = {\mit\Phi}(z,\theta^A)$.

The expansion of the right-hand side of \re{O} in powers of $z_j-z_k$ yields
an infinite tower of operators
\be
\prod_{l=1}^L (\partial_{z_l})^{k_l} \, \mathbb{O}(Z_1,\ldots,Z_L) \bigg|_{z_l=0}
= \tr \left\{ D_+^{k_1} {\mit\Phi} (0, \theta^A_1) \ldots D_+^{k_L} {\mit\Phi}
(0, \theta^A_L) \right\} \, , \label{Wilson}
\ee
where we took into account that $n \cdot D = D_+ = \partial_+$ in the gauge $A_+
(x)=0$. Expanding further the right-hand side of \re{Wilson} in powers of
$\theta^A-$variables, one obtains a set of composite operators built from
propagating fields. Notice that not all of them are Wilson operators. The reason
for this is that the superfields, Eqs.~\re{N=4-field}, \re{N=2-field},
\re{N=1-field} and \re{N=0-field}, involve nonlocal operators $\partial_+^{-1}
A(x)$ and $\partial_+^{-1}\bar\lambda_A(x)$, while the Wilson operators can only
involve $\partial_+ A(x)$, $\bar\lambda_A(x)$ and their derivatives. For
instance, expansion of the light-cone operator $\mathbb{O}(Z_1,Z_2)$ in the
$\mathcal{N}=4$ theory gives rise to the following dimension two operators:
$\phi^{AB} \bar\phi_{AB}(0)$, $\partial_+^{-1} A\,
\partial_+ \bar A(0)$ and $\partial_+^{-1} \bar \lambda_A\, \lambda_A(0)$. Among
them only the first one is a Wilson operator. Although naively one might expect
that this operator could mix under renormalization with the other two, the
locality of the $\mathcal{N}=4$ theory prohibits such mixing.

To eliminate the contribution of nonlocal operators to \re{O}, one has to
project the nonlocal operator \re{O} onto a ``physical'' subspace of {\sl
local} Wilson operators
\be
\mathbb{O}^{\scriptscriptstyle \rm W}(Z_1,\ldots,Z_L) = \Pi \cdot
\,\mathbb{O}(Z_1,\ldots,Z_L) \equiv \tr \left\{ {\mit\Phi}^{\scriptscriptstyle
\rm W} (z_1, \theta^A_1) \ldots {\mit\Phi}^{\scriptscriptstyle \rm W} (z_L,
\theta^A_L) \right\} \, , \label{projector}
\ee
where the notation was introduced for the superfield
\be
{\mit\Phi}^{\scriptscriptstyle \rm W}(z, \theta^A) = {\mit\Phi}(z , \theta^A )
- {\mit\Phi}(0,0) - Z \cdot\partial_Z {\mit\Phi}(0,0) \, , \label{Phi-W}
\ee
with $Z \cdot\partial_Z \equiv z \partial_z + \theta^A \partial_{\theta^A}$.
The superfield ${\mit\Phi}^{\scriptscriptstyle \rm W} (z, \theta^A)$
does not involve nonlocal gauge and fermion operators, $\partial_+^{-1}
A(0)$, $A(0)$ and $\partial_+^{-1}\bar\lambda^A(0)$, and, therefore,
$\mathbb{O}^{\scriptscriptstyle \rm W}(Z_1,...,Z_L)$ generates only
Wilson operators. It is easy to verify using \re{projector} that $\Pi$ is a
projector, $\Pi^2 = \Pi$.

The Wilson operators generated by the light-cone operator \re{projector} mix
under renormalization. To find their anomalous dimensions one has to
diagonalize the corresponding mixing matrix. For the light-cone SYM theories
introduced in the previous section, this matrix can be calculated in
perturbation theory using the supergraph Feynman technique developed in
Ref.~\cite{BriLinNil83}. The size of the mixing matrix depends on the total
number of derivatives in \re{Wilson} and it rapidly grows as this number
increases. To avoid such complication and in order to reveal a hidden symmetry
of the evolution equations, it is convenient to work directly with nonlocal
light-cone operators \re{O}. Having established the scale dependence of the
light-cone operator \re{O}, one can reconstruct the mixing matrix by
substituting the nonlocal operator $\mathbb{O}(Z_1,Z_2,\ldots,Z_L)$ by its
expansion in terms of local Wilson operators \re{Wilson}.

\begin{figure}[t]
\psfrag{k}[cc][cc]{${}_k$}\psfrag{k+1}[cc][cc]{${}_{k+1}$}
\centerline{\epsfxsize14.0cm\epsfbox{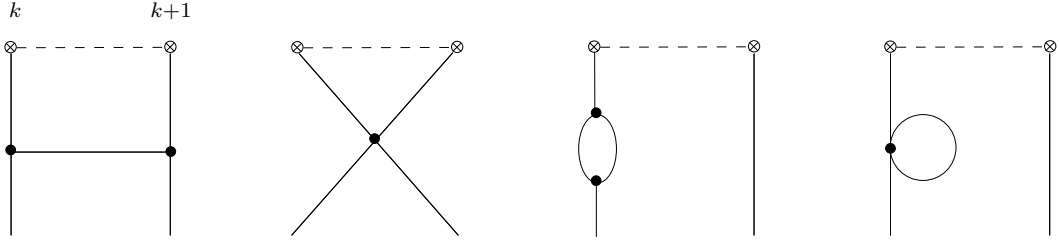}} \vspace*{0.3cm}
\caption[]{Feynman supergraphs defining one-loop contribution to the two-particle
evolution kernel $\mathbb{H}_{k,k+1}$ in the multi-color limit. The dashed line
denotes the `+'--direction on the light-cone and the open circles define the
position of the superfields.}%
\label{fig1}%
\end{figure}

In the multi-color limit, $N_c \to \infty$ and $g^2 N_c=\rm fixed$, to the lowest
order of perturbation theory, quantum corrections to \re{O} arise due to
interaction between two nearest neighbor superfields ${\mit\Phi}(Z_k)$ and
${\mit\Phi}(Z_{k+1})$ (with $k=1,\ldots,L$). The corresponding Feynman diagrams
are displayed in Fig.\ \ref{fig1}. They involve cubic and quartic vertices which
can be read off the light-cone actions, Eqs.~\re{N=4-S}, \re{N=2-S}, \re{N=1-S}
and \re{N=0-S}. The first two diagrams are divergent due to light-cone separation
of the superfields while the remaining two diagrams contain conventional
ultraviolet divergences due to the renormalization of the superfields. To
one-loop order the renormalization group (Callan-Symanzik) equation for the
nonlocal operator \re{O} can be written as~\cite{BFKL,BB}
\be
\left\{
\mu\frac{\partial}{\partial\mu}
+
\beta(g)
\frac{\partial}{\partial g} \right\} \mathbb{O}(Z_1,Z_2,\ldots,Z_L)
=
-
\frac{g^2 N_c}{8\pi^2} \left[ \mathbb{H}\cdot
\mathbb{O}\right](Z_1,Z_2,\ldots,Z_L) + \mathcal{O} \left((g^2 N_c)^2\right)
\, ,
\label{EQ}
\ee
where $\mathbb{H}$ is some integral operator and $\beta (g) = 0$ in the ${\cal N}
= 4$ SYM theory. This equation is just a Ward identity for the dilatation
operator in the ${\cal N}-$extended SYM theory in the multi-color limit (see,
e.g., \cite{BraKorMul}). The evolution kernel $\mathbb{H}$ defines the
representation of the dilatation operator on the space of light-cone superfields
\re{O}. Its eigenvalues determine the spectrum of anomalous dimensions of Wilson
operators in the $\mathcal{N}-$extended SYM theory.

The possible form of the evolution kernel $\mathbb{H}$ is constrained by the
symmetries of the underlying SYM theory.
%
%
To one-loop order, in the multi-color limit the operator $\mathbb{H}$ can be
written as the sum over two-particle nearest neighbor interactions
\be
\mathbb{H} = \mathbb{H}_{12} +  \ldots + \mathbb{H}_{L-1,L} +
\mathbb{H}_{L,1}\,. \label{H}
\ee
The two-particle kernel $\mathbb{H}_{k,k+1}$ receives contribution from the
Feynman diagrams shown in Fig.~\ref{fig1}. We will establish the explicit form of
$\mathbb{H}$ in the $\mathcal{N}-$extended SYM theory in Sect.~5.

Secondly, we recall that $\mathbb{H}$ acts on the product of superfields \re{O}
whose expansion around $Z_k=0$ generates both Wilson operators and nonlocal
operators. The OPE implies that the Wilson operators cannot mix with nonlocal
operators while the inverse is possible. Thus, the evolution kernel $\mathbb{H}$
in \re{EQ} has to transform ``physical'' operators
$\mathbb{O}^{\scriptscriptstyle \rm W}(Z_1,\ldots,Z_L)$, Eq.~\re{projector}, into
themselves. To satisfy this requirement it suffices to demand that $\Pi \,
\mathbb{H} \, (1 - \Pi) = 0$ where the projector $\Pi$ was introduced in Eq.\
\re{projector}. We shall check this relation in Sect.~5 (see Eq.~\re{H-property}
below).

Finally, the third constraint is imposed by the invariance of the classical
action of the light-cone SYM theory, Eqs.~\re{N=4-S}, \re{N=2-S}, \re{N=1-S}
and \re{N=0-S}, under superconformal transformations. This symmetry survives at
the quantum level in the $\mathcal{N}=4$ theory, while at $\mathcal{N}=2$,
$\mathcal{N}=1$ and $\mathcal{N}=0$ it is broken by quantum corrections. In the
latter case, however, symmetry breaking effects affect the evolution equations
\re{EQ} starting only from the two-loop level~\cite{Mul94}. This means that the
one-loop operator $\mathbb{H}$ possesses the symmetry of the classical action
and, therefore, it has to commute with the generators of the
$\mathcal{N}-$extended superconformal group. As we will show in the next
section, this imposes severe restrictions on the possible form of the evolution
kernel~\re{H}.

\section{Superconformal symmetry on the light-cone}
\label{Sect4}

The ${\cal N}-$extended superconformal algebra $SU(2,2|\mathcal{N})$ contains 15
even charges ${\mathbf P}_\mu$, ${\mathbf M}_{\mu\nu}$, ${\mathbf D}$ and
${\mathbf K}_\mu$ and $4 {\cal N}$ odd charges ${\mathbf Q}_{\alpha A}$,
$\mathbf{\bar { Q}}^{\dot\alpha A}$, ${\mathbf S}_\alpha^A$,
$\mathbf{\bar{S}}^{\dot\alpha}_A$ which are two-component Weyl spinors and
${\scriptstyle A} = 1, 2, \dots, {\cal N}$. It also contains additional bosonic
chiral charge ${\mathbf R}$ and, in case of extended ${\cal N} > 1$
supersymmetries, the charges ${\mathbf T}_A{}^B$ satisfying the $SU({\cal N})$
commutation relations~\cite{Sohn}. Scalar light-cone superfield ${\mit\Phi}
(x, \theta^A , \bar\theta_A)$ realizes a representation of this algebra. Its
infinitesimal variations under superconformal transformations look like
\begin{equation}
\delta^G {\mit\Phi} (x, \theta^A , \bar\theta_A)
=
i [ {\mit\Phi} (x , \theta^A , \bar\theta_A), {\mathbf G} ]
=
- i G \, {\mit\Phi} (x, \theta^A , \bar\theta_A)
\, ,
\label{trans}
\end{equation}
where ${\mathbf G} = \varepsilon^\mu{\mathbf P}_\mu ,\, \varepsilon^{\mu\nu}
{\mathbf M}_{\mu\nu} , \dots$ for even generators and ${\mathbf G} = \xi
{\mathbf Q} , \dots$ for odd generators with $\xi$ being a constant
Grassmann-valued Weyl spinor. In this representation, the quantum-field
operator ${\mathbf G}$ is represented by the operator $G$ acting on even and
odd coordinates of the superfield~\cite{Sohn}.

Let us examine the action of superconformal algebra on the operators \re{O}. We
remind that the superfields entering \re{O} are located on the light-cone along
the `$+$'-direction defined by the light-like vector $n_\mu$. In order to
preserve this property, we have to restrict ourselves to superconformal
transformations that map the `$+$'-direction on the light-cone into itself. Under
these restrictions the full superconformal algebra is reduced to its subalgebra,
the so-called collinear $SL(2|\mathcal{N})$ superalgebra, containing the
following generators
\begin{equation}
{\mathbf P}_+
\, , \quad
{\mathbf M}_{-+}
\, , \quad
{\mathbf D}
\, , \quad
{\mathbf K}_-
\, , \quad
\mathbf{Q}_{+ \alpha A}
\, , \quad
\mathbf{\bar Q}_{+}^{\dot\alpha A}
\, , \quad
{\mathbf S}_{- \alpha}^A
\, , \quad
\mathbf{\bar S}_{- A}^{\dot\alpha}
\, , \quad
{\mathbf R}
\, , \quad
{\mathbf T}_A{}^B
\, , \quad
{\mathbf M}_{12}
\, ,
\label{collinear}
\end{equation}
where for the odd charges the $+/-$ subscript indicates ``good''/``bad''
components of the Weyl spinor.%
\footnote{For arbitrary Weyl spinors $\chi_\alpha$ and $\bar\chi^{\dot\alpha}$
the projection onto the ``good'' and ``bad'' components looks like ${\chi}_{\pm
\alpha} =\ft12
\bar\sigma^\mp{}_{\alpha\dot\beta} \, \sigma^{\pm \; \dot\beta\gamma}
{\chi}_\gamma$ and ${\bar \chi}_\pm^{\dot\alpha}\equiv \ft12 \sigma^{\mp \;
\dot\alpha\beta} \,
\bar\sigma^\pm{}_{\beta\dot\gamma}
{\bar \chi}^{\dot\gamma}$.} In the light-cone formalism, such one-component
spinors can be described by a complex Grassmann field without any Lorentz
index.

In the light-cone formalism, the action of the generators of the collinear
superalgebra \re{collinear} on the chiral scalar superfield ${\mit\Phi}
(x n_\mu, \theta^A , \bar\theta_A)$ can be represented as differential
operators acting on $z=x + \ft12 \, \bar\theta \cdot \theta$ and $\theta^A$
while the remaining generators of the superconformal algebra have a nonlinear
realization on the superfield~\cite{BriTol83}. Introducing linear combinations
of the operators $i P_+ \equiv - L^-$, $\ft12 i K_- \equiv L^+$, $\ft12 i
( D + M_{-+} ) \equiv L^0$, $i ( D - M_{-+} ) \equiv E$, $i Q_{+ A} \equiv
\sqrt[4]{8} \, V_{A}^-$, $i {\bar Q}{}_{+}^A \equiv - i \sqrt[4]{8} \,
{W}^{A,-}$, $i   S_-^{A} \equiv - \sqrt[4]{32} \, W^{A,+}$, $i  {\bar S}{}_{-A}
\equiv i \sqrt[4]{32} \, {V}_{A}^+$ and $\ft14 (1 - \ft{4}{{\cal N}})
R + \ft12   M_{12} \equiv  B$, we find that they admit the following
representation for the chiral scalar superfield
\begin{equation}
\label{sl2}
\begin{array}{llll}
{L}^- = -\partial_z \, , \ & {L}^+ = 2 j\, z + z^2\partial_z + z \left(
\theta\cdot \partial_\theta \right) \, , \ & {L}^0 = j + z
\partial_z + \ft12\left( \theta\cdot \partial_\theta \right)
\, , \ & {E}= t \, , \\ [3mm] {W}{}^{A,-} = \theta^A \, \partial_z \, , \ &
{W}{}^{A,+} = \theta^A [ 2j  +  z \partial_z +  \left( \theta\cdot
\partial_\theta \right) ] \, , \ & {V}^-_{A} = \partial_{\theta^A} \, , \ &
{V}^+_{A} = z\partial_{\theta^A} \, , \\ [3mm] \, \ & {T}_B{}^A = \theta^A
\partial_{\theta^B} - \ft1{\mathcal{N}} \, \delta_B^A \left( \theta\cdot
\partial_\theta \right) \, , \ & {B} = - j - \ft12 \left( 1 - \ft{2}{{\cal N}}
\right) \left( \theta\cdot \partial_\theta \right) \, . \ &
\end{array}
\end{equation}
where $\partial_z \equiv \partial/\partial z$ and $\theta \cdot \partial_\theta
\equiv \theta^A \partial/\partial\theta^A$. The operators \re{sl2} satisfy the
$SL(2|\mathcal{N})$ (anti-) commutation relations~\cite{FraSorSci}. In
Eq.~\re{sl2}, $j = \ft12 (s + \ell)$ is the {\sl conformal spin} and $t = \ell -
s$ is the {\sl twist} of the superfield~\cite{BraKorMul}. Also, $\ell$ and $s$
are correspondingly the canonical dimension and projection of the spin on the
`$+$'- direction of the superfield defined as $i [ \mathbf{D}, {\mit\Phi}(0,0,0)]
= \ell \, {\mit\Phi}(0,0,0)$ and $i [ \mathbf{M}_{-+},{\mit\Phi}(0,0,0)] = s \,
{\mit\Phi}(0,0,0)$. For the scalar superfields defined in \re{N=4-field},
\re{N=2-field}, \re{N=1-field} and \re{N=0-field} one has $\ell=0$ and $s=-1$
leading to
\be
j = - \frac12 \, , \qquad t = 1 \, .
\label{j=-1/2}
\ee

The one-loop evolution kernel $\mathbb{H}$ has to respect the superconformal
symmetry and, therefore, it has to commute with the generators of the
$SL(2|\mathcal{N})$ collinear superalgebra. Let us determine a general form of
the operator satisfying these conditions. To begin with, we consider the
evolution of the light-cone operator \re{O} in the special limit when all
odd variables are set equal to zero,
\be
\mathbb{O}^{(0)}(x_1,\ldots,x_L) \equiv
\mathbb{O}(Z_1,\ldots,Z_L)\bigg|_{\theta_1^A=\ldots=\theta^A_L=0}= \tr \left\{
{\mit\Phi}(x_1 n_\mu,0) \ldots {\mit\Phi}(x_L n_\mu,0) \right\}.
\label{O-sl2}
\ee
Since ${\mit\Phi}(xn_\mu,0)=\partial_x^{-1} A(xn_\mu)$, the operator \re{O-sl2}
is reduced to the product of $L$ gauge fields of helicity $+1$. To one-loop
order, the operator \re{O-sl2} can only mix with the operators containing the
same number of superfields and the same total helicity. This means that
$\mathbb{O}^{(0)}(x_1,\ldots,x_L)$ evolves autonomously under renormalization
group transformations and, therefore, the evolution kernel $\mathbb{H}$ has to
map light-cone operators \re{O-sl2} into themselves. Notice that for the
operators \re{O-sl2} the collinear superconformal group \re{sl2} is reduced to
its $SL(2)$ subgroup with the generators ${l}^- = -\sum_k \partial_{x_k}$, ${l}^+
= \sum_k 2 j\, x_k + x_k^2\partial_{x_k}$ and ${l}^0 = \sum_k j +
x_k\partial_{x_k}$. Therefore, the evolution kernel governing the scale
dependence of the operators \re{O-sl2} has to be $SL(2)$ invariant.

A general form of the $SL(2)$ invariant kernel has been found in
Ref.~\cite{BraDerKorMan99}. To one-loop order, in the multi-color limit it has
the form \re{H} with the two-particle kernel given by
\be
[\mathbb{H}_{12}\cdot \mathbb{O}^{(0)}](x_1,\ldots,x_L) = \int_0^1
d\alpha\int_0^{\bar\alpha} d\beta\,
(\alpha\beta)^{2j-2}\varphi\left(\frac{\bar\alpha\bar\beta}{\alpha\beta}
\right) \mathbb{O}^{(0)} (x_1-\bar\alpha x_{12},x_2+\bar\beta
x_{12},x_3,\ldots,x_L)\,, \label{H-sl2}
\ee
where $\bar\alpha=1-\alpha$, $\bar\beta=1-\beta$ and $x_{12}=x_1-x_2$. Here $j$
is the conformal spin of the fields entering
\re{O-sl2} and $\varphi$ is an arbitrary function.%
\footnote{Strictly speaking, this formula holds only for $j \ge 1/2$ while for $j
< 1/2$ it has to be modified to make the integral convergent (see below).} The
operator \re{H-sl2} has a simple interpretation---acting on the product of $L$
fields situated on the light-cone it displays only two of them (labeled as $1$
and $2$) in the direction of each other. The explicit form of the function
$\varphi(\xi)$ is not fixed by the $SL(2)$ invariance. It depends both on the
underlying gauge theory and the operator under
consideration~\cite{BraDerKorMan99}.

The two-particle kernel $\mathbb{H}_{12}$ has to be invariant under the
superconformal transformations and act locally on the superfields $\Phi(Z_1)$ and
$\Phi(Z_2)$. This implies in particular that
$[\mathbb{H}_{12},V^{\pm,{\scriptscriptstyle (1)}}_{A}+V^{\pm,{\scriptscriptstyle
(2)}}_{A}]=0$ and $[\mathbb{H}_{12},V^{\pm,{\scriptscriptstyle (k)}}_{A}]=0$
where $k\ge 3$ and the superscript $(k)$ indicates that the charges  $V^\pm_A$
defined in \re{sl2} act on the coordinates of $k$th field. These charges generate
shifts in the superspaces along the odd directions
\be
\mathbb{O}^{(0)}(x_1,x_2) = {\rm e}^{ - \epsilon^A {}[ V^{-,{\scriptscriptstyle
(1)}}_{A} + V^{-,{\scriptscriptstyle (2)}}_{A} {}] - \chi^A {}[
V^{+,{\scriptscriptstyle (1)}}_{A} + V^{+,{\scriptscriptstyle (2)}}_{A} {}]}
\,\mathbb{O}(Z_1,Z_2) \, ,
\label{rotation}
\ee
where $\chi^A=(\theta_1^A-\theta_2^A)/x_{12}$ and
$\epsilon^A=(x_1\theta_2^A-x_2\theta_1^A)/x_{12}$.
Combining together \re{rotation} and \re{H-sl2} one gets
\be
[\mathbb{H}_{12}\cdot \mathbb{O}](Z_1,\ldots,Z_L) = \int_0^1
d\alpha\int_0^{\bar\alpha} d\beta\,
(\alpha\beta)^{2j-2}\varphi\left(\frac{\bar\alpha\bar\beta}{\alpha\beta}
\right) \mathbb{O}(Z_1-\bar\alpha Z_{12},Z_2+\bar\beta Z_{12},Z_3,\ldots,Z_L)
\, , \label{H-sl2N}
\ee
where $Z_{12} = Z_1-Z_2\equiv (x_1-x_2,\theta_1^1-\theta_2^1,\ldots,
\theta_1^\mathcal{N}-\theta_2^\mathcal{N})$.
As before, the superconformal symmetry does not allow one to fix the explicit
form of the weight function $\varphi(\xi)$.

To summarize, Eq.~\re{H-sl2N} defines the most general form of the two-particle
evolution kernel consistent with the symmetries of $\mathcal{N}-$extended SYM
theory. This kernel has a transparent interpretation in the superspace. Acting
on the light-cone operator \re{O} it displaces the superfields located at the
points $Z_1$ and $Z_2$ in the direction of each other. As we will show in the
next section, the fact that both even and odd coordinates of the
superfields are modified simultaneously implies certain pattern of the mixing
between Wilson operators.

\section{One-loop dilatation operator}
\label{Sect5}

The expression for the two-particle evolution kernel \re{H-sl2N}
depends on the conformal spin of the superfield $j=-1/2$, Eq.~\re{j=-1/2}, and
yet unknown function $\varphi(\xi)$. To determine this function one has to
calculate the Feynman supergraphs shown in Fig.~\ref{fig1} and match their
divergent part into a general expression for the two-particle kernel
$\mathbb{H}_{12}$, Eq.~\re{H-sl2N}.

Going through the calculation of Feynman supergraphs we find that in the
$\mathcal{N}=4$, $\mathcal{N}=2$, $\mathcal{N}=1$ and $\mathcal{N}=0$ SYM
theories the one-loop two-particle evolution kernel $\mathbb{H}_{12}$ has the
{\sl same}, universal form. Namely, $\mathbb{H}_{12}$ is factorized into a
product of two commuting operators
\be
\mathbb{H}_{12}  = \mathbb{V}_{12} \, (1-\Pi_{12})\,,
\label{H12-int}
\ee
with $[\mathbb{V}_{12},\Pi_{12}]=0$. The operator $\mathbb{V}_{12}$ is given by
\re{H-sl2N} for $j=-1/2$ and $\varphi(\xi)=-\delta(\xi)$
\ba
&&\mathbb{V}_{12} \mathbb{O} (Z_1,...,Z_L) = \int_0^1
\frac{d\alpha}{(1-\alpha)\alpha^2} \, \bigg\{ 2 \alpha^2\, \mathbb{O}
(Z_1,Z_2,...,Z_L)
\label{V-int}
\\
&& \hspace*{+30mm} - \mathbb{O}(\alpha Z_1+(1-\alpha) Z_{2},Z_2,...,Z_L) -
\mathbb{O}(Z_1,\alpha Z_2+(1-\alpha) Z_{1},...,Z_L) \bigg\} \nonumber \, . \ \ \
\ \
\ea
The second operator is a projector, $(\Pi_{12})^2=\Pi_{12}$. It is defined as
\be
\Pi_{12} \mathbb{O} (Z_1,...,Z_L) = \ft12\left(1+ Z_{12}\cdot
\partial_{Z} \right)\mathbb{O}(Z,Z_2,...,Z_L)\bigg|_{Z=Z_2}
+ \ft12\left(1+ Z_{21}\cdot
\partial_{Z}\right)\mathbb{O}(Z_1,Z,...,Z_L)\bigg|_{Z=Z_1},
\label{Pi12}
\ee
where $(Z_{12}\cdot \partial_{Z})\equiv (z_1-z_2)\partial_{z}
+(\theta_{1}^A-\theta_{2}^A) \partial_{\theta^A}$. Examining \re{V-int} we see
that the integral over $\alpha$ is divergent for $\alpha\to 0$. Using \re{V-int}
and \re{Pi12} one can check that divergences cancel in the expression for
$[\mathbb{V}_{12}(1- \Pi_{12}) \cdot\mathbb{O}] (Z_1,\ldots,Z_L)$ and, therefore,
the integral operator \re{H12-int} is well-defined. We would like to stress that
Eqs.~\re{H12-int} -- \re{Pi12} are valid only for the light-cone operators \re{O}
built from the chiral superfields in the $\mathcal{N}-$extended SYM.  The
$\mathcal{N}-$dependence enters into Eqs.~\re{H12-int} -- \re{Pi12} entirely
through the dimension of the superspace
$Z=(x,\theta^1,\ldots,\theta^\mathcal{N})$. This means that in order to go, for
example, from  $\mathcal{N}=4$ down to $\mathcal{N}=2$ one has to put
$\theta^3=\theta^4=0$.

It is straightforward to verify that the operators $\mathbb{V}_{12}$ and
$\Pi_{12}$ commute with the generators of the superconformal algebra \re{sl2} and
the same is true for $\mathbb{H}_{12}$. The appearance of the projector
$1-\Pi_{12}$ in the right-hand side of \re{H12-int} can be understood as follows.
It can be deduced from the evolution equation \re{EQ} that, because of this
projector, the operator $\Pi_{12}\mathbb{O} (Z_1,Z_2)$ has a vanishing anomalous
dimension. Indeed, according to the definition \re{Pi12}, the operator $\Pi_{12}
\mathbb{O} (Z_1,Z_2)$ contains bilinear products of nonlocal fields,
$\partial_+^{-1}A(0)$, $A(0)$ and $\partial_+^{-1}\bar\lambda(0)$, which do not
contain ultraviolet divergences as long as the Mandelstam-Leibbrandt prescription
is used.

The evolution kernel defined in Eqs.~\re{H12-int} and \re{H} governs the scale
dependence of a nonlocal light-cone operator $\mathbb{O}(Z_1,\ldots,Z_L)$,
Eq.~\re{O}. As was already mentioned, $\mathbb{O}(Z_1,\ldots,Z_L)$ is a
generating function for both the Wilson operators and spurious nonlocal
operators. The latter operators can be removed by applying the projector $\Pi$,
Eq.~\re{projector}, to both sides of the evolution equation \re{EQ}. One verifies
using \re{H12-int} and \re{Pi12} that the operators $\mathbb{H}_{12}$ and
$\Pi_{12}$ satisfy the following relations
\be
\Pi \mathbb{H}_{12} (1-\Pi) = 0\,,\qquad \Pi \Pi_{12} = 0\,.
\label{H-property}
\ee
Multiplying both sides of \re{EQ} by $\Pi$ one finds  that the ``physical''
light-cone operator $\mathbb{O}^{\scriptscriptstyle \rm W}(Z_1,...,Z_L)$,
Eq.~\re{projector}, evolves autonomously with the corresponding two-particle
evolution kernel given by
\be
\mathbb{H}_{12}^{\scriptscriptstyle \rm W} = \Pi \mathbb{H}_{12} = \Pi
\mathbb{H}_{12} \Pi = \Pi \mathbb{V}_{12}(1-\Pi_{12}) \Pi  = \Pi (1-\Pi_{12})
\mathbb{V}_{12} \Pi = \Pi \mathbb{V}_{12} \Pi\,.
\label{H-W}
\ee

Combined together Eqs.~\re{H}, \re{H12-int}, \re{V-int} and \re{Pi12} define the
one-loop evolution kernel for the light-cone operators \re{O} in $\mathcal{N}=4$,
$\mathcal{N}=2$, $\mathcal{N}=1$ and $\mathcal{N}=0$ SYM theories in the
multi-color limit. The origin of such universality is the following. According to
\re{H-sl2}, $\varphi(\xi)$ determines the two-particle kernel for maximal
helicity operators \re{O-sl2}. This kernel describes the RG evolution of the
light-cone operator $\partial_{x_1}^{-1}A(x_1 n_\mu) \,
\partial^{-1}_{x_2} A(x_2 n_\mu)$ built from two helicity $+1$ gauge fields. To
one-loop order, the corresponding Feynman diagrams involve cubic and quartic pure
gluonic vertices and, therefore, they are only sensitive to the pure gluonic part
of the $\mathcal{N}$-extended SYM action. The latter is the same for the
light-cone actions \re{N=4-S}, \re{N=2-S}, \re{N=1-S} and \re{N=0-S}.

Expanding both sides of \re{EQ} in powers of even and odd variables, $z_k$ and
$\theta^A_k$, respectively, one can obtain from \re{H12-int} the mixing matrix
for Wilson operators of the maximal Lorentz spin in the $\mathcal{N}-$extended
SYM. To illustrate the predictive power of \re{H12-int} let us derive the mixing
matrices for three different sets of the Wilson operators: maximal helicity gauge
field operators, scalar operators  in the $\mathcal{N}=4$ SYM theory with no
derivatives and twist-two $SO(6)$ singlet operators involving an arbitrary number
of derivatives. Obviously, these three examples do not cover all possible Wilson
operators. A general classification of the solutions to \re{EQ} will be given
elsewhere.
\medskip

\noindent (i) \underline{\sl Maximal helicity Wilson gauge operators} are
defined in the light-cone gauge $A_+(x)=0$ as
\be
O_{j_1j_2\ldots j_L}^{\rm h}(0) = \tr\{ \partial_+^{j_1+1}
A(0)\,\partial_+^{j_2+1} A(0)\ldots \partial_+^{j_L+1} A(0)\}\,,
\label{max-hel}
\ee
with $0\le j_k < \infty$ counting the number of derivatives. To one-loop order,
in the multi-color limit, the operators \re{max-hel} mix under renormalization
with the operators $O^{\rm h}_{j_1'j_2'\ldots j_L'}(0)$ having the same total
number of derivatives $j_1'+\ldots+j_L'$. According to \re{N=4-field}, the
helicity $+1$ gauge field determines the lowest component of the $\mathcal{N} =
4$ superfield leading to $\partial_+^{j+1} A(x)=\partial_{x}^{j+2} {\mit\Phi}
(x,0,0)$. Therefore, applying $\partial_{z_1}^{j_1+2}\ldots
\partial_{z_L}^{j_L+2}$ to both sides of \re{EQ} and putting
$Z_1=\ldots=Z_L=0$ afterwards, one obtains from \re{EQ} the evolution equation
for the Wilson operators \re{max-hel}. The corresponding mixing matrix takes the
form (see Eq.~\re{H})
\be
[\mathbb{H}]_{j_1\ldots j_L}^{j_1'\ldots j_L'} =
V{}_{j_1j_2}^{j_1'j_2'}\delta_{j_3}^{j_3'} \ldots\delta_{j_L}^{j_L'}+\cdots
+\delta_{j_1}^{j_1'} \delta_{j_2}^{j_2'}\ldots V{}_{j_{L-1}j_L}^{j_{L-1}'j_L'}+
\delta_{j_2}^{j_2'}\delta_{j_3}^{j_3'} \ldots V{}_{j_{L}j_1}^{j_{L}'j_1'}\,,
\label{H-scalar}
\ee
where the two-particle mixing matrix $V{}_{j_1j_2}^{j_1'j_2'}$ describes the
transition $\partial_+^{j_1+1}\!\! A\,\partial_+^{j_2+1}\!\!  A \to
\partial_+^{j_1'+1}\!\!  A\,\partial_+^{j_2'+1}\!\!  A$. It is related to the
two-particle evolution kernel as
\be
\partial_{z_1}^{j_1+2}
\partial_{z_2}^{j_2+2}\left[\mathbb{H}_{12} \cdot
\mathbb{O}\right](Z_1,Z_2)\bigg|_{Z_1=Z_2=0}=\sum_{ j_1'+j_2'=j_1+j_2}
V{}_{j_1j_2}^{j_1'j_2'} \ \partial_+^{j_1'+1}\! A(0)\,\partial_+^{j_2'+1}
\!A(0)
\, .
\label{VV}
\ee
This relation establishes the correspondence between the mixing matrix of local
Wilson operators and evolution kernels of nonlocal light-cone operators. Making
use of \re{H12-int}, the left-hand side of \re{VV} can be written after some
algebra as
\be
\mbox{Eq.$\re{VV}=$}\int_0^1\frac{d\alpha}{\bar\alpha} \alpha^2
\left[2\partial_{z_1}^{j_1}\partial_{z_2}^{j_2} -\alpha^{j_1}
(\bar\alpha\partial_{z_1}\!+\partial_{z_2})^{j_2}\partial_{z_1}^{j_1}-\alpha^{j_2}
(\bar\alpha\partial_{z_2}\!+\partial_{z_1})^{j_1}\partial_{z_2}^{j_2}
\right]\partial_{z_1} A(z_1)\partial_{z_2} A(z_2)\bigg|_{z_k=0}\,.
\ee
This expression serves as a generating function for the matrix elements
$V{}_{j_1j_2}^{j_1'j_2'}$ entering \re{VV}. The mixing matrix \re{H-scalar} for
the maximal helicity operators \re{max-hel} with an arbitrary number of
derivatives and a given number of fields $L$ has an infinite size. In addition,
as was mentioned in Sect.~1, it possesses a hidden symmetry. Namely, the
evolution kernel \re{H-scalar} can be mapped into a Hamiltonian of completely
integrable Heisenberg $SL(2)$ spin chain of length $L$.

\medskip

\noindent (ii) \underline{\sl Composite scalar operators} are built from six real
scalar fields $\phi_j(x)$ ($j=1,\ldots,6$)
\be
O_{j_1j_2\ldots j_L} = \tr \{ \phi_{j_1}(0) \phi_{j_2}(0) \ldots \phi_{j_L}(0)
\} \, , \qquad \phi_j(x) = \ft1{2\sqrt{2}}\, {\mit\Sigma}_j^{AB} \bar
\phi_{AB}(x) \, , \label{real-scalar}
\ee
which are given by linear combinations of $\bar\phi_{AB}$. Here, $j = 1,
\ldots, 6$ and ${\mit\Sigma}_j^{AB}$ are the chiral blocks of Dirac matrices in
six-dimensional Euclidean space \cite{BelDerKorMan03}. In the multi-color
limit, to one-loop order the scalar operators $O_{j_1 \ldots j_L}(0)$ mix with
each other under renormalization. Similar to the previous case, one uses the
relation between the scalar fields and the $\mathcal{N}=4$ superfield
\re{N=4-field}, $\phi_j(x) = \ft{i}{2\sqrt{2}} {\mit\Sigma}_j^{AB}
\partial_{\theta^A}\partial_{\theta^B} {\mit\Phi}(x, \theta^A , 0)|_{\theta^A =
0}$, to derive from \re{EQ} the evolution equation for the scalar operators
\re{real-scalar}. The corresponding mixing matrix takes the same form as before,
Eq.~\re{H-scalar},  but the expression for the two-particle mixing matrix
$V{}_{j_1j_2}^{j_1'j_2'}$ describing the transition $\phi_{j_1} \phi_{j_2} \to
\phi_{j_1'} \phi_{j_2'}$ is different
\be
- \ft18 \left( {\mit\Sigma}_{j_1}^{AB} \partial_{\theta_1^A}
\partial_{\theta_1^B} \right) \left( {\mit\Sigma}_{j_2}^{CD}
\partial_{\theta_2^C} \partial_{\theta_2^D} \right) \left[ \mathbb{H}_{12} \cdot
\mathbb{O} \right] (Z_1, Z_2) \bigg|_{Z_1 = Z_2 = 0} = \sum_{j_1'j_2'}
V{}_{j_1j_2}^{j_1'j_2'} \ \phi_{j_1'}(0) \phi_{j_2'}(0) \, .
\ee
Substituting \re{H12-int} into this relation one finds after some algebra
\be
V{}_{j_1j_2}^{j_1'j_2'} = \delta_{j_1}^{j_1'} \delta_{j_2}^{j_2'} + \ft12\,
\delta_{j_1j_2}\delta^{j_1'j_2'}- \delta_{j_1}^{j_2'} \delta_{j_2}^{j_1'}  =
 \bigg|\ \bigg| +
\ft12\, {{\bigcup} \atop {\bigcap}}-\mbox{$\bigg\backslash\hspace*{-4.2mm}
\bigg\slash$} \, . \label{V}
\ee
The expressions \re{H-scalar} and \re{V} define the one-loop evolution kernel for
the scalar operators \re{real-scalar} in the multi-color limit. They are in
agreement with the results of Ref.~\cite{MinZar02}. The mixing matrix
$[\mathbb{H}]_{j_1\ldots j_L}^{j_1'\ldots j_L'}$ for the product of $L$ scalar
operators has a finite dimension $6^L$. As in the previous case, it has a hidden
symmetry -- this matrix can be mapped into a Hamiltonian of completely integrable
Heisenberg $SO(6)$ spin chain of length $L$ \cite{MinZar02}.

\medskip

\noindent (iii) \underline{\sl Twist-two $SO(6)$ singlet parity-even gauge
operators} are defined as
\be
O^g (N) = \tr \{ F_{+ \mu} D_+^N F_{+}{}^\mu \} =
 \tr
\{
\partial_+^{N+1}\bar A\,\partial_+ A
+
\partial_+ \bar A \,\partial_+^{N+1} A
\}
\, , \label{tw-2-g}
\ee
where in the second relation we adopted the light-cone gauge $A_+(x) = 0$.
They mix under renormalization with the twist-two fermion and scalar
operators of the same canonical dimension
\be
O^q (N) = \tr \{
\partial_+^{N+1} \bar\lambda_A \lambda^A-\bar\lambda_A \partial_+^{N+1} \lambda^A
\} \, , \qquad O^s (N) = \tr\{ \bar\phi_{AB} \partial_+^{N+2} \phi^{AB} \} \, ,
\label{tw-2-s}
\ee
as well as with the operators containing total derivatives $\partial_+^n O^a (N -
n)$, with $a = g,q,s$ and $n = 1, \ldots, N$. The latter operators can be
effectively eliminated by going over to the forward matrix elements of the
operators \re{tw-2-g} and \re{tw-2-s}. The mixing between the operators
\re{tw-2-g} and \re{tw-2-s} is described by a $3\times 3$ matrix of the anomalous
dimensions $\gamma_{ab}(N)$ (with $a,b=g,q,s$). To calculate its one-loop
expression from \re{EQ}, one uses the relations $
\partial_+ A(x) = \partial_x^2\, {\mit\Phi}(x,0,0)$
and
$\partial_+ \bar A(x) = - \textrm{d}_\theta\,{\mit\Phi}(x,0,0)$, where
the $\mathcal{N}=4$ superfield is given by \re{N=4-field} and
$\textrm{d}_\theta\equiv \ft1{4!} \varepsilon^{ABCD}
\partial_{\theta^A}\partial_{\theta^B}\partial_{\theta^C}\partial_{\theta^D}$.
Then, the evolution equation for the gauge field operator \re{tw-2-g} can be
derived from \re{EQ} as
\be
\mu\frac{d}{d\mu}\vev{O^g(N)}= \frac{g^2
N_c}{8\pi^2}\vev{\left(\partial_{z_1}^N+\partial_{z_2}^N\right)
\textrm{d}_{\theta_1}
\partial_{z_2}^2 \left[2\mathbb{H}_{12}
\cdot \mathbb{O}\right](Z_1,Z_2)}\bigg|_{Z_k=0}= \frac{g^2
N_c}{8\pi^2}\sum_{b=g,q,s} \gamma_{g b}(N)\,\vev{O^b(N)}\,, \label{tw-2-kernel}
\ee
where $\vev{...}$ stands for the forward matrix element and the additional factor
of 2 takes into account that the evolution kernel for the two-particle operators
equals $\mathbb{H}=\mathbb{H}_{12} + \mathbb{H}_{21} = 2 \mathbb{H}_{12}$.
Substituting \re{H12-int} into \re{tw-2-kernel} one finds after some algebra
\ba
&& \gamma_{gg}(N) = 4\left[
\psi(1)-\psi(N+1)-\frac2{N+2}+\frac1{N+3}-\frac1{N+4}\right]\,, \nonumber
\\
&& \gamma_{gq}(N) = \frac4{N+1} - \frac4{N+2}  +  \frac2{N+3} \,,\qquad
\gamma_{gs}(N) = \frac{2}{(N+1)(N+2)}\,,
\ea
where $\psi(x)=d\ln \Gamma(x)/dx$ is the Euler function. The remaining entries of
the matrix of anomalous dimensions can be calculated in a similar manner. This
leads to the expression for $\gamma_{ab}(N)$ which is in agreement with the
results of Ref.~\cite{KotLip02}. As we will show in the next section,
$\gamma_{ab}(N)$ can be mapped into a two-particle Hamiltonian of the Heisenberg
$SL(2|4)$ spin chain.

\section{Heisenberg $SL(2|\mathcal{N})$ spin chain}
\label{Sect6}

Let us show that the one-loop evolution kernel for light-cone operators \re{O} in
the $\mathcal{N}-$extended SYM theory defined in \re{H} and \re{H12-int}
possesses a hidden integrability -- it can be identified as a Hamiltonian of the
Heisenberg $SL(2| \mathcal{N})$ spin chain. To this end, we shall construct the
$R-$matrix on the space of light-cone superfields and argue that its logarithmic
derivative coincides with the expression for the two-particle evolution kernel
\re{H-W}.

The light-cone scalar superfield ${\mit\Phi}(z,{\theta}^A)$ defines a
representation of the superconformal $SL(2|\mathcal{N})$ algebra that we shall
denote as $V$. The fact that ${\mit\Phi}(z,{\theta}^A)$ is chiral implies that
$V$ is the so-called atypical representation~\cite{FraSorSci}. The generators of
the algebra are realized on $V$ as differential operators \re{sl2}. The states
$\{1,z,\theta^A,z\theta^A,\theta^A\theta^B,\ldots\}\in V$ define coefficient
functions in the expansion of the superfield around the origin in the superspace
${\mit\Phi}(z,\theta^A)= {\mit\Phi}(0,0) + Z\cdot\partial_Z {\mit\Phi}(0,0)+
\ft1{2!}(Z\cdot\partial_Z)^2{\mit\Phi}(0,0)+\ldots$ with $1$ being the lowest
weight in $V$. The quadratic Casimir operator is~\cite{FraSorSci} \footnote{The
singularity of $\mathbb{J}^2$ at $\mathcal{N}=2$ is spurious since it can be
removed by adding to the r.h.s.\ of \re{Casimir} an infinite c-number
correction.}
\be
\mathbb{J}^2 = (L^0)^2 + L^+ L^- + (\mathcal{N}-1) L^0
+\frac{\mathcal{N}}{\mathcal{N}-2} B^2 - V^+_A W^{A,-} - W^+_A V^{A,-}-\ft12\,
T^B{}_A T^A{}_B\,.
\label{Casimir}
\ee
For the superfield ${\mit\Phi}(z,\theta^A)$ one has $ \mathbb{J}^2\,
{\mit\Phi}(z,{\theta}^A) = j\left[j+
\mathcal{N}-1+\frac{\mathcal{N}j}{\mathcal{N}-2}
\right]{\mit\Phi}(z,{\theta}^A)\,, $ with $j=-1/2$ being the conformal spin of
the superfield \re{j=-1/2}.

A distinguished feature of $V$ is that for $j=-1/2$ it contains an invariant
subspace spanned by the vectors $V_0 =
\{1,z,\theta^1,\ldots,\theta^\mathcal{N}\}$. The corresponding superfield
${\mit\Phi}_0(z,\theta^A)= {\mit\Phi}(0,0) + z\partial_z {\mit\Phi}(0,0)+
\theta^A\partial_{\theta^A}{\mit\Phi}(0,0)$ is built entirely from spurious
components of fields and is annihilated by the projector \re{projector},
$\Pi{\mit\Phi}_0(z,\theta^A)=0$. Thus, the ``physical'' superfield
${\mit\Phi}^{\scriptscriptstyle \rm W}={\mit\Phi}(z,\theta^A)-{\mit\Phi}_0
(z,\theta^A)$ defined in Eq.~\re{Phi-W} belongs to the quotient of the two spaces
$V_{\scriptscriptstyle \rm W} = V/V_0$. Let us now consider the product of two
superfields ${\mit\Phi}(Z_1){\mit\Phi}(Z_2)$ belonging to the tensor product
$V\otimes V$. The generators of the superconformal algebra are given on $V\otimes
V$ by the sum of the differential operators \re{sl2} acting on the $Z_1-$ and
$Z_2-$coordinates. Using \re{Casimir} one can define the corresponding
two-particle Casimir operator $\mathbb{J}_{12}^2$ and realize it on $V\otimes V$
as a $2\times 2$ matrix $[\mathbb{J}_{12}^2]_{11} =\Pi \mathbb{J}_{12}^2\Pi$,
$[\mathbb{J}_{12}^2]_{12}=\Pi \mathbb{J}_{12}^2 (1-\Pi)$, etc. Since $(1-\Pi)
V\otimes V =V_0\otimes V_0 - V_0\otimes V - V\otimes V_0$ and $V_0$ is the
invariant subspace annihilated by the projector $\Pi$, one has
$[\mathbb{J}_{12}^2]_{12}=0$, or equivalently,
\be
\Pi \,\mathbb{J}_{12}^2 (1-\Pi) \ {\mit\Phi}(Z_1){\mit\Phi}(Z_2) =0\,.
\label{L2-Pi}
\ee
This relation implies that the Casimir operator is given a triangular matrix
$\mathbb{J}_{12}^2=\bigl(\begin{smallmatrix}
* & 0 \\ * & *
\end{smallmatrix}\bigr)$.
The same is true for an arbitrary two-particle operator like $\mathbb{H}_{12}$
(see Eq.~\re{H-property}) invariant under the superconformal transformations.

To reveal integrable structures of the evolution equations \re{EQ} in the
$\mathcal{N}-$extended SYM theory, we apply the $R-$matrix approach~\cite{QISM}.
As the starting point, we introduce the Lax operator for the $SL(2|\mathcal{N})$
algebra~\cite{Kulish,Frahm,DKK}. It is given by the graded matrix of dimension
$\mathcal{N}+2$ whose entries are linear combinations of the generators of the
superconformal algebra \re{sl2}
\be
[\mathbb{L} (u)]_{ab}=\left(
\begin{array}{ccc}
u + L^0 + \frac{{\cal N}}{{\cal N} - 2} B & -W^{B, -} & L^-
\\ [3mm]
- V^+_A & u \delta_A^B - T_A{}^B + \frac{2}{{\cal N} - 2} B \delta_A^B & V^-_A
\\ [3mm]
L^+ & -W^{B, +} & u - L^0 + \frac{{\cal N}}{{\cal N} - 2} B
\end{array}
\right) ,
\label{Lax}
\ee
where $u$ is a complex spectral parameter and the indices run over
$a=(0,{\scriptstyle A}, {\cal N} + 1)$ and $b=(0, {\scriptstyle B}, {\cal N} +
1)$ with ${\scriptstyle A, B}=1,\ldots,\mathcal{N}$.

Let us now define an integral operator $\mathbb{R}_{12}(u)$ acting on the tensor
product $V\otimes V$ as
\ba
[\mathbb{R}_{12}(u)\cdot \mathbb{O}](Z_1,Z_2)=
\frac{u\sin(\pi u)}{\pi} \int_{0}^{1} d \alpha\int_{0}^{1-\alpha} d \beta\,
(\alpha\beta)^{-u-1} (1-\alpha-\beta)^{2j+u-1} &&
\nonumber \\[3mm]
&& \hspace*{-70mm} \times\, \mathbb{O}(\alpha Z_1+(1-\alpha)Z_2, \beta
Z_2+(1-\beta)Z_1)\,,
\label{R-matrix}
\ea
where $j=-1/2$ and $u$ is a spectral parameter. One can verify that the operator
\re{R-matrix} satisfies the $SL(2|\mathcal{N})$ Yang-Baxter equation
\be
\mathbb{R}_{12}(u) \mathbb{R}_{13}(u+v) \mathbb{R}_{23}(v)=
\mathbb{R}_{23}(v) \mathbb{R}_{13}(u+v) \mathbb{R}_{12}(u)
\label{YB}
\ee
and interchanges the Lax operators $\mathbb{L}_1(u)= \mathbb{L}(u) \otimes \II$
and $\mathbb{L}_2(u)= \II \otimes \mathbb{L}(u)$
\be
{\mathbb{L}}_1(u)\,{\mathbb{L}}_2(u+v)\, \mathbb{R}_{12}(v) =
\mathbb{R}_{12}(v)\, {\mathbb{L}}_2(u+v)\, \mathbb{L}_1(u)\,.
\label{LLR}
\ee
Here, the Lax operators $\mathbb{L}_k(u)$ (with $k=1,2$) are given by \re{Lax}
with the superconformal charges acting on the $Z_k-$coordinates.
If one puts in \re{R-matrix} the odd coordinates equal to zero,
$Z_k=(z_k,\theta^A_k=0)$, the operator $\mathbb{R}_{12}(u)$ is reduced to the
known expression for the $SL(2)$ invariant $R-$matrix acting along the $z-$axis
in the superspace~\cite{DKK}.

The relation \re{R-matrix} can be seen as a special case of the
$SL(2|\mathcal{N})$ invariant operator \re{H-sl2N} for $\varphi(\xi)=
(\xi-1)^{2j+u-1}{u\sin(\pi u)}/{\pi}$. The operator $\mathbb{R}_{12}(u)$
satisfies the same relation \re{L2-Pi} as the two-particle Casimir operator
$\mathbb{J}_{12}^2$ and, therefore, it is given by a triangular matrix
\be
\mathbb{R}_{12}(u)=\left(%
\begin{array}{cc}
\mathbb{R}_{12}^{\scriptscriptstyle \rm W}(u)  & 0 \\
\ast & \ast \\
\end{array}%
\right),\qquad \mathbb{R}_{12}^{\scriptscriptstyle \rm W}(u)\equiv
\Pi\,\mathbb{R}_{12}(u) =\Pi\,\mathbb{R}_{12}(u)\,\Pi\,.
\ee
Substituting this expression into \re{YB} one finds that
$\mathbb{R}_{12}^{\scriptscriptstyle \rm W}(u)$ satisfies the Yang-Baxter
equation \re{YB}. Thus, the operator $\mathbb{R}_{12}^{\scriptscriptstyle \rm
W}(u)$ is the $R-$matrix on the tensor product of ``physical'' spaces
$V_{\scriptscriptstyle \rm W}\otimes V_{\scriptscriptstyle \rm W}$ with
$V_{\scriptscriptstyle \rm W}\equiv V\backslash V_0$. Invoking the standard
arguments \cite{QISM}, one finds that its logarithmic derivative at $u=0$ defines
the two-particle Hamiltonian of a completely integrable, homogenous $
\textrm{XXX}$ Heisenberg $SL(2|\mathcal{N})$ spin chain
\be
\mathbb{H}_{12}^{\scriptscriptstyle \rm XXX} = \frac{d
}{du}\ln\mathbb{R}_{12}^{\scriptscriptstyle \rm W}(u)\bigg|_{u=0}=\Pi
\left(\frac{d }{du}\ln\mathbb{R}_{12}(u)\bigg|_{u=0}\right)\Pi\,.
\label{H12-Heisenberg}
\ee
To calculate the derivative entering this expression, one expands the integral in
the right-hand side of \re{R-matrix} at small $u$ and projects the result onto
$\Pi$, Eq.~\re{projector}. In this way, one finds after some algebra
\be
\mathbb{R}_{12}^{\scriptscriptstyle \rm W}(u) = \mathbb{P}_{12}\left[1 + u\cdot
\Pi \mathbb{V}_{12} \Pi + \mathcal{O}(u^2)\right]\,,
\ee
where the integral operator $\mathbb{V}_{12}$ was defined in \re{V-int} and
$\mathbb{P}_{12}$ is the permutation operator, $\mathbb{P}_{12}
\mathbb{O}(Z_1,Z_2)= \mathbb{O}(Z_2,Z_1)$. Substituting this relation into
\re{H12-Heisenberg} one recovers the two-particle evolution kernel \re{H-W},
$\mathbb{H}_{12}^{\scriptscriptstyle \rm XXX}=\mathbb{H}_{12}^{\scriptscriptstyle
\rm W}$. This allows us to identify the evolution kernel for the ``physical''
light-cone operators $\mathbb{O}^{\scriptscriptstyle \rm W}(Z_1,\ldots,Z_L)$ as
the Hamiltonian of a {\sl noncompact} Heisenberg $SL(2|\mathcal{N})$ spin
chain of length $L$ and the single-particle spin $j=-1/2$. Similar integrable
spin chains have been previously studied in Refs.~\cite{DKK}.

To understand the properties of the obtained expression for the $R-$matrix one
chooses $j=-\ft12+\epsilon$ and examines Eqs.~\re{R-matrix} and
\re{H12-Heisenberg} in the limit $\epsilon\to 0$. For $\epsilon\neq 0$, the
$SL(2|\mathcal{N})$ generators are given by \re{sl2}. The corresponding
representation space $V(\epsilon)$ is irreducible for $\epsilon\neq 0$  while at
$\epsilon=0$ one has $V(0)=V_0\oplus V_{\scriptscriptstyle \rm W}$. The tensor
product $V(\epsilon)\otimes V(\epsilon)$ can be decomposed into irreducible
components with the lowest weights $\Psi_l(Z_1,Z_2)$ ($l=0,1,\ldots$)
\be
\Psi_0=1\,,\qquad \Psi_{k}=\theta_{12}^{A_1}\ldots \theta_{12}^{A_k} \,,\qquad
\Psi_{n+\mathcal{N}}=\varepsilon_{A_1\ldots
A_{\mathcal{N}}}\theta_{12}^{A_1}\ldots \theta_{12}^{A_\mathcal{N}}z_{12}^n\,,
\ee
where $\theta_{12}^A= \theta_1^A - \theta_2^A$, $z_{12}=z_1-z_2$, $1\le k \le
\mathcal{N}-1$ and $0 \le n < \infty$. These states satisfy the relations $L^-
\Psi_l = W^{A,-} \Psi_l = V^-_A \Psi_l=0$ and diagonalize the two-particle
Casimir operator \re{Casimir}
\be
\left(\mathbb{J}_{12}^2-\Delta_j\right)\Psi_l = (l+2j)(l+2j-1) \Psi_l = J_{12}
(J_{12}-1) \Psi_l\,,
\ee
with $\Delta_j=2j\mathcal{N} \left[ 1+ \frac{2j}{\mathcal{N}-2}\right]$. Here the
notation was introduced for the two-particle superconformal spin
$J_{12}=l+2j=-1+l+2\epsilon$ with $l=0,1,...$. Substituting the lowest weights
into \re{R-matrix}, one evaluates the eigenvalues of the $R-$matrix
\be
\mathbb{R}_{12}(u) \Psi_l =
(-1)^l\frac{\Gamma(1-u)\Gamma(J_{12}+u)}{\Gamma(1+u)\Gamma(J_{12}-u)}\Psi_l\,.
\ee
For the modules with the lowest weights $\Psi_0$ and $\Psi_1$ the eigenvalues of
$[d\ln \mathbb{R}_{12}(u)/du]|_{u=0}$ behave as $\sim 1/\epsilon$ while for the
remaining modules $\Psi_l$ ($l\ge 2$) it approaches a finite value as $\epsilon
\to 0$. Notice that the projector $\Pi$ annihilates the modules $\Psi_0$ and
$\Psi_1$ and, therefore, the Hamiltonian \re{H12-Heisenberg} is well-defined for
$\epsilon\to 0$
\be
\mathbb{H}_{12}^{^{\rm W}} \Psi_{l+2}^{^{\rm W}} = 2\left[\psi(l+1)
-\psi(1)\right]\Psi_{l+2}^{^{\rm W}}\,, \label{H-eig}
\ee
where $\Psi_{l}^{^{\rm W}} =\Pi\Psi_{l}$ and $\Psi_{0}^{^{\rm W}} =
\Psi_{1}^{^{\rm W}}=0$. Equation \re{H-eig} defines the eigenvalues of the
two-particle evolution kernel \re{H-W} as a function of the superconformal spin
$J_{12}=l-1$.

The Hamiltonian of the Heisenberg $SL(2|\mathcal{N})$ spin magnet possesses
a set of the integrals of motion. In the $R-$matrix approach, one can find their
explicit form by constructing the auxiliary transfer matrix. It is equal to the
supertrace of the product of the Lax operators \re{Lax} and is given by a
polynomial in $u$ of degree $L$ with operator-valued coefficients,
\be
t_L(u) = {\rm str} \left\{\mathbb{L}_L(u) \ldots \mathbb{L}_2(u) \mathbb{L}_1(u)
\right\} = (2-\mathcal{N}) u^L + {q}_2 u^{L-2} + \ldots + q_L\,,
\ee
where $q_2= \mathbb{J}_{12\ldots L}^2 - L[\ft12\Delta_{-1/2}+\ft34]$ is related
to the total superconformal spin. It follows from the Yang-Baxter equations
\re{YB} and \re{LLR} that the operators $q_k$ commute among themselves and with
the evolution kernel $\mathbb{H}$. The spectral problem for the Heisenberg
$SL(2|\mathcal{N})$ spin magnet can be solved within the Bethe Ansatz (see, e.g.,
\cite{Saleur}). Using its eigenspectrum one can construct the basis of
superconformal operators having an autonomous scale dependence and evaluate the
corresponding anomalous dimensions. We shall return to this problem in a
forthcoming publication.

\section{Conclusions}
\label{Sect7}

In this paper, we have demonstrated that the one-loop evolution kernel governing
the scale dependence of the single trace product of chiral light-cone superfields
in the $\mathcal{N}-$extended SYM theory coincides in the multi-color limit with
the Hamiltonian of the Heisenberg $SL(2|\mathcal{N})$ spin chain. We constructed
the evolution kernel as an integral operator acting on the superspace and found
that it has the same, universal form in the $\mathcal{N}=0$, $\mathcal{N}=1$,
$\mathcal{N}=2$ and $\mathcal{N}=4$ SYM theories. The only difference is that
going over from the $\mathcal{N}=0$ to the maximally symmetric $\mathcal{N}=4$
theory, one has to increase the number of odd dimensions in the superspace.

We already mentioned that in multi-color QCD the one-loop dilatation operator
acting on the Wilson operators of maximal helicity built from the field strength
tensor and sandwiched between the quark fields in the fundamental representation
coincides with the Hamiltonian of the open Heisenberg $SL(2)$ spin chain. It is
straightforward to lift this structure into the superspace and define the
corresponding $SL(2|\mathcal{N})$ invariant dilatation operator. It is natural to
expect that it describes the scale dependence of the product of the light-cone
superfields in the $\mathcal{N}-$extended SYM theory supplemented with the matter
in the fundamental representation.

In our analysis we restricted ourselves to Wilson operators of the maximal
Lorentz spin built from the ``good'' components of the fields. We recall that
in the light-cone formalism, supersymmetry is realized differently for the
``good'' and ``bad'' components. In our consideration we made use of a part
of superconformal generators which admit a linear realization on the
superspace. Making use of the remaining part of superconformal generators,
one can extend the analysis to the Wilson operators of lower spin and those
built from ``bad'' components.

The dilatation operator constructed in this paper acts on the space of a single
trace product of chiral superfields. In the $\mathcal{N}=4$ SYM it covers all
possible Wilson operators of the maximal Lorentz spin while in the SYM theory
with less supersymmetry it should be supplemented by mixed products of chiral and
antichiral light-cone superfields. In the latter case, the dilatation operator
can be realized as a Hamiltonian of the spin chain but its integrability property
will be lost. As was shown in Refs.~\cite{BraDerKorMan99,Bel99}, the additional
terms in the Hamiltonian responsible for breaking the integrability lead to the
formation of a {\sl mass gap}\ in the spectrum of the anomalous dimensions in
multi-color QCD. This issue deserves further investigation in the
$\mathcal{N}-$extended SYM theory.

\bigskip

We would like to thank  V.~Braun, A.~Gorsky and A.~Vainshtein for interesting
discussions. The present work was supported in part by the US Department of
Energy under contract DE-FG02-93ER40762 (A.B.), by Sofya Kovalevskaya programme
of Alexander von Humboldt Foundation (A.M.), by the NATO Fellowship (A.M.\
and S.D.) and in part by the grant 00-01-005-00 from the Russian Foundation for
Fundamental Research (A.M.\ and S.D.).



\end{document}